\documentclass{elsart}
\usepackage{natbib}

\usepackage{epsfig}

\def\bc{\begin{center}}
\def\ec{\end{center}}
\def\emi{\end{minipage}}
\def\epi{\end{picture}}
\def\bearray{\begin{eqnarray}}
\def\eearray{\end{eqnarray}}

\newcommand{\be}{\begin{equation}}
\newcommand{\ee}{\end{equation}}
\newcommand\ccb{{c\bar{c}}}

\begin{document}
\begin{flushright}
ECT*--02--03\\
February 2002\\
hep-ph/0201066\\[5em]
\end{flushright}

\begin{frontmatter}

\title{Constraints on quark energy loss \\ from Drell-Yan data}

\author{Fran\c{c}ois Arleo\thanksref{email}}
\address{ECT* and INFN, G.C. di Trento,\\
Strada delle Tabarelle, 286\\
38050 Villazzano (Trento), Italy}

\thanks[email]{{\it Email address:} \texttt{francois@ect.it} (Fran\c{c}ois Arleo)}

\begin{abstract}
A leading-order analysis of E866/NuSea and NA3 Drell-Yan data in nuclei is carried out. At Fermilab energy, the large uncertainties in the amount of sea quark shadowing prohibit clarifying the origin of the nuclear dependence observed experimentally. On the other hand, the small shadowing contribution to the Drell-Yan process in $\pi^-$--$A$ collisions at SPS allows one to set tight constraints on the energy loss of fast quarks in nuclear matter. We find the transport coefficient to be $\hat{q} = 0.24 \pm 0.18$~GeV/fm$^2$ that corresponds to a mean energy loss per unit length $- dE / dz = 0.20 \pm 0.15$~GeV/fm for $E_q >$~50~GeV quarks in a large ($A \approx 200$) nucleus.
\end{abstract}

\end{frontmatter}

%%%%%%%%%%%%%%%%%%%%%%%%%%%%%%%%%%%%%%%%%%%%%%%%%%%%%%%%%%%%%%%%%%%%
\section{Introduction}
%%%%%%%%%%%%%%%%%%%%%%%%%%%%%%%%%%%%%%%%%%%%%%%%%%%%%%%%%%%%%%%%%%%%

Energy loss of hard partons in hot QCD matter is expected to be large~\cite{GP1,Baier1}. Consequently, it has been suggested that the depletion of high $p_\perp$ jets (jet quenching) in heavy ion collisions may be a signal for quark-gluon plasma formation~\cite{Wang,BDMPS}. Recently, a lot of excitement was created as PHENIX data revealed that high $p_\perp$ hadron spectra were found to be substantially suppressed in the most central Au-Au collisions at RHIC with respect to the extrapolation from $p$--$p$ data~\cite{phenix}. Multiple scattering of a high energy parton traversing a large nucleus (``cold'' QCD matter) has been studied similarly by Baier et al. (BDMPS) in Ref.~\cite{BDMPS} in which a numerical estimate for the expected parton mean energy loss per unit length $-dE / dz$ is given. 

The Drell-Yan mechanism is a process particularly suited for the study of quark energy loss in nuclei as the lepton pair does not strongly interact with the surrounding medium. Furthermore, new $p$--$A$ data recently became available from the E866/NuSea experiment at Fermilab~\cite{e866}. Subsequently, two recent attempts to extract the quantity $-dE / dz$ from these data have been carried out~\cite{e866,johnson}. However, their results do not agree as the amount of sea quark shadowing assumed in both analysis strongly differs. As we shall see later, the poorly known shadowing corrections at Fermilab energy indeed makes a model-independent extraction of quark energy loss unlikely.

In this letter, we discuss constraints on $-dE / dz$ from the analysis of both E866/NuSea~\cite{e866} and NA3~\cite{na3} Drell-Yan dimuon data in hadron nucleus reactions. The procedure followed is detailed in Section~\ref{sect:dy} after having given the leading-order Drell-Yan production cross section in nuclei. The results given in Section~\ref{sect:results} are discussed and compared with previous studies in the last section.

%%%%%%%%%%%%%%%%%%%%%%%%%%%%%%%%%%%%%%%%%%%%%%%%%%%%%%%%%%%%%%%%%%%%
\section{Nuclear dependence of Drell-Yan production}\label{sect:dy}
\subsubsection*{Leading order production cross section}
To leading order (LO) in perturbation theory, the Drell-Yan (DY) process describes dilepton production from quark-antiquark annihilation. The invariant mass $M$ of the lepton pair is set by the center-of-mass energy of the $q\bar{q}$ collision $\sqrt{\hat{s}} = (x_1\, x_2\, s)^{1/2}$ where $x_1$ (resp. $x_2$) is the momentum fraction carried by the beam (resp. target) parton and $\sqrt{s}$ is the center-of-mass energy of the hadronic collision. The differential partonic cross section $q\bar{q}\to l^+ l^-$ has been computed to leading order in~\cite{gavin} and is given by
\be\label{eq:partonic_xs}
\frac{{\rm d}\hat{\sigma}}{{\rm d}M}=\frac{8 \pi \alpha^2}{9 M} \,e_q^2\, \delta(\hat{s} - M^2).
\ee
The hadronic DY production cross section is then obtained from the convolution of the partonic cross section~(\ref{eq:partonic_xs}) with the quark distributions in the beam and in the target hadron, evaluated at $x_1$ and $x_2$ respectively, and at a factorization scale $\mu^2 = M^2$, i.e.,
\be
\frac{{\rm d}\sigma (h h')}{{\rm d}x_1\,{\rm d}M} = \frac{8 \pi \alpha^2}{9 M} \,\frac{1}{x_1 s} \,\sum_q \,e_q^2\, \left( f_q^h(x_1) f_{\bar{q}}^{h'}(x_2) + f_{\bar{q}}^h(x_1) f_q^{h'}(x_2) \right)
\ee
where the sum is carried out over the light quark sector $q = u, d, s$ and $x_2 = M^2 / x_1 s$ after integration over the delta function\footnote{In the following, parton distributions will always be evaluated at the hard scale $\mu^2 = M^2$. For simplicity, we shall drop the explicit dependence and use $f_i(x)=f_i(x,\mu^2=M^2)$ in the notations.}. Looking at the $x_1$ dependence of DY production, we shall rather deal in the following with the single differential cross section
\be\label{eq:dyxs}
\frac{{\rm d}\sigma (h h')}{{\rm d}x_1} = \frac{8 \pi \alpha^2}{9 \,x_1\, s} \sum_q e_q^2\,\int \frac{{\rm d}M}{M} \left( f_q^h(x_1) f_{\bar{q}}^{h'}(x_2) + f_{\bar{q}}^h(x_1) f_q^{h'}(x_2) \right)
\ee
where the integration over the dilepton mass is performed in the range between the $\ccb$ and the $b\bar{b}$ resonances. The LO cross section (\ref{eq:dyxs}) proves to describe the whole trend of DY data within a so-called $K\sim 2$ factor which might be attributed to large next-to-leading order (NLO) corrections. Since we are primarily interested here in the nuclear dependence of Drell-Yan production, one may reasonably expect these higher-order corrections to cancel in the production {\it ratio}
\be\label{eq:ratio}
R^h(A/B,x_1) = \frac{B}{A}\,\left(\frac{{\rm d}\sigma(h A)}{{\rm d}x_1}\right) \times \left(\frac{{\rm d}\sigma(h B)}{{\rm d}x_1}\right)^{-1}
\ee
in a heavy (A) over a light (B) nucleus. We shall therefore restrict ourselves to a LO analysis throughout this paper. In the absence of nuclear effects, the production cross section in hadron-nucleus reactions $\sigma(h A)$ appearing in~(\ref{eq:ratio}) is given by
\begin{eqnarray}\label{eq:dyxsA}
\frac{{\rm d}\sigma (h A)}{{\rm d}x_1} = \frac{8 \pi \alpha^2}{9 \,x_1\, s} \sum_q e_q^2\,\int \frac{{\rm d}M}{M} \biggr[ &&Z \left(f_q^h(x_1) f_{\bar{q}}^{p}(x_2) + f_{\bar{q}}^h(x_1) f_q^{p}(x_2)\right) \\
& & + (A-Z) \left(f_q^h(x_1) f_{\bar{q}}^{n}(x_2) + f_{\bar{q}}^h(x_1) f_q^{n}(x_2) \right) \biggr]\nonumber
\end{eqnarray}
after separating into terms involving protons and neutrons in the target nucleus. Isospin effects will remain small as long as the parton densities in the proton and in the neutron do not differ strongly $f_i^p(x_2,\mu^2) \approx f_i^n(x_2,\mu^2)$, that is, at small $x_2 = M^2 / x_1 s \ll 1$. In this particular case, DY production scales with the atomic mass number $A$, hence the nuclear production ratio (\ref{eq:ratio}) is equal to one. 

Beyond isospin corrections, several nuclear effects --- such as shadowing or parton energy loss --- might affect the Drell-Yan process and lead to an unusual $A$-dependence in dilepton production. Let us discuss now how these mechanisms modify the nuclear production cross section~(\ref{eq:dyxsA}).

\subsubsection*{Shadowing}
Nuclear deep inelastic scattering (DIS) data indicate that parton distributions in nuclei differ significantly from those in a proton~\cite{arneodo}. In particular, a significant depletion (``shadowing'') of high momentum ($0.3 < x_2 < 0.7$) as well as low momentum ($x_2 < 0.05$) partons in large nuclei has been reported experimentally. The origin of nuclear shadowing is still rather unclear. It may be attributed to the multiple scattering of the struck quark with the target, which can be removed into effective nuclear parton distribution functions according to the QCD factorization theorem~\cite{QS}. Therefore, Drell-Yan dilepton production off nuclei can be estimated using the nuclear densities $f_i^{p / A}$ (resp. $f_i^{n / A}$) instead of the ``free'' parton distributions $f_i^p$ (resp. $f_i^n$) in Eq.~(\ref{eq:dyxsA}). In the following numerical applications, we will assume that the nuclear parton distributions $f_i^{p / A}$ factorize into a nuclear contribution $R_i^A$ and the parton distribution in a proton $f_i^p$, i.e.,
\be\label{eq:paramEKS98}
f_i^{p / A} (x, \mu^2) = R_i^A (x, \mu^2) \times f_i^p (x, \mu^2)
\ee 
where the function $R_i^A (x, \mu^2)$ has been parameterized by Eskola, Kolhinen, and Salgado (EKS98) from a leading-order DGLAP analysis of DIS data~\cite{EKS98}. It is worth pointing out that E772 Drell-Yan data~\cite{e772} have also been taken into account in their fitting procedure to further constrain sea quark shadowing in the intermediate $x_2$ range ($x_2 \sim 0.1$)~\cite{EKS98}. We shall come back to this observation when discussing results in Sections~\ref{sect:results} and~\ref{sect:discussion}.

\subsubsection*{Parton energy loss}

The quark (antiquark) from the projectile may scatter through the nucleus before the hard $q\bar{q}$ annihilation process occurs. The medium induced gluon emission from the incoming parton with energy $E_p$ leads to a radiative parton energy loss $\epsilon$. In the BDMPS approach, the distribution $D(\epsilon)$ in the energy loss is characterized by a typical energy scale $\omega_c$ proportional to the square of the length $L$ of traversed nuclear matter
\be\label{eq:omc}
\omega_c \,=\, \frac{1}{2}\,\hat{q}\,L^2.
\ee
The so-called ``transport coefficient'' $\hat{q}$ relates the $p_\perp$ broadening of the parton to the length $L$~\cite{BDMPS}. It is expected to depend on both the small~$x$ gluon distribution and the density $\rho$ of scattering centers of the medium (here, $\rho = 0.15 \,{\rm fm}^{-3}$). Assuming the hard process to take place uniformly in the nucleus, the length $L$ is proportional to the nuclear radius, $L = 3/4\, R$. 

Neglecting interference effects in the multiple gluon radiation, Baier et al. give the distribution $D(\epsilon)$ a simple integral representation~\cite{BDMS_quench}. For the applications to come, the distribution $D(\epsilon)$ has been computed numerically in the soft gluon approximation, $\epsilon \sim \omega_c \ll E_p$, from Eq.~(18) of Ref.~\cite{BDMS_quench}. The {\it mean} BDMPS energy loss $\Delta E$ of the incoming quark is given by~\cite{BDMPS}
\begin{eqnarray}\label{eq:de}
- \Delta E \,\equiv\, \int d\epsilon\,\epsilon\,D(\epsilon) = \frac{1}{2}\,\alpha_S\, C_R \,\omega_c \,\propto\, L^2
\end{eqnarray}
with $C_R$ being the color charge of the parton ($C_R = 4/3$ for quarks) and $\alpha_S = 1/2$ the strong coupling constant. In the following, we shall therefore write the mean energy loss per unit length as
\be\label{eq:eloss}
- \frac{d E}{d z} \,\equiv\, - \frac{\Delta E}{L} \,=\, \delta\,\times\,\left(\frac{L}{10\,{\rm fm}}\right)
\ee
where $\delta$ is a free parameter simply related to the transport coefficient $\hat{q}$ through (\ref{eq:omc}) and (\ref{eq:de}). 

This multiple scattering effect shifts the quark (antiquark) momentum fraction from $x_1 + \Delta x_1(\epsilon)$ to $x_1$ at the point of fusion, with
\be\label{eq:shift}
\Delta x_1(\epsilon) = \frac{\epsilon}{E_h}
\ee
and where $E_h$ is the projectile hadron energy in the nucleus rest frame\footnote{Let us drop in the following the explicit $\epsilon$ dependence of $\Delta x_1$ for clarity.}. Consequently, the parton densities $f_i^h(x)$ have to be evaluated at $(x_1+\Delta x_1)$ in the nuclear production cross section~(\ref{eq:dyxsA}). Because of the steep behavior of the valence quark distributions at large $x_1$ (e.g., $u_v \sim (1-x_1)^{3 - 4}$ in a proton), even a small shift $\Delta x_1$ may substantially suppress Drell-Yan production in a large nucleus as compared to a light one. We further note that the larger $x_1$, the stronger the suppression $u_v(x_1+\Delta x_1)/u_v(x_1)$.

\subsubsection*{Analysis of Drell-Yan data}

In the previous sections we have stressed that nuclear mechanisms affect the Drell-Yan process. In the most general case, the production cross section in hadron-nucleus reactions will read
\begin{eqnarray}\label{eq:dyxsA_general}
\frac{{\rm d}\sigma (h A)}{{\rm d}x_1}  & = & \frac{8 \pi \alpha^2}{9 x_1 s} \sum_q e_q^2\int \frac{{\rm d}M}{M} \int {\rm d}\epsilon \,D(\epsilon)\, \\
&&\biggr[ Z f_q^h(x_1 +\Delta x_1) f_{\bar{q}}^{p / A}(x_2) 
 + (A-Z) f_q^h(x_1 +\Delta x_1) f_{\bar{q}}^{n / A}(x_2) \nonumber\\
& & + Z f_{\bar{q}}^h(x_1 +\Delta x_1) f_q^{p / A}(x_2) + (A-Z) f_{\bar{q}}^h(x_1 +\Delta x_1) f_q^{n / A}(x_2) \biggr], \nonumber
\end{eqnarray}
taking both nuclear shadowing and parton energy loss effects into account. At large $x_1$, the restricted phase space ($\epsilon < (1-x_1)\,E_h$) makes the effects of quenching even more pronounced. In the limit of no shadowing ($f_i^{p / A} = f_i^p$) and vanishing energy loss ($\Delta  x_1 = 0$), one retrieves the usual production cross section~(\ref{eq:dyxsA}).

The aim of our study is to investigate whether the above mentioned effects manifest themselves in available data, and, if so, to possibly disentangle shadowing from energy loss contributions to the nuclear dependence of Drell-Yan production. In particular, it would be most interesting to set some constraints on the amount of parton energy loss in nuclear matter. To achieve such a goal, a close comparison between data and theory has been carried out. Using Eq.~(\ref{eq:dyxsA_general}), Drell-Yan production is computed with the following four options:
\begin{enumerate}
\item[(i)]  $f_i^{p / A} = f_i^{p}$, $\Delta x_1 = 0$
\item[(ii)] $f_i^{p / A} \ne f_i^{p}$, $\Delta x_1 = 0$
\item[(iii)]$f_i^{p / A} = f_i^{p}$, $\Delta x_1 \ne 0$
\item[(iv)] $f_i^{p / A} \ne f_i^{p}$, $\Delta x_1 \ne 0$.
\end{enumerate}
No nuclear effect is assumed in the first set (i), while shadowing and energy loss~corrections are considered in turn ((ii) and (iii) respectively). Both effects are then combined in the last case~(iv). Shadowing corrections were taken from the EKS98 parameterization~(Eq.~(\ref{eq:paramEKS98})) and the momentum fraction shift $\Delta x_1$ is given by Eq.~(\ref{eq:shift}) with $\delta$ (and hence the coefficient $\hat{q}$) kept as a free parameter fitted to the data. We made use of the MRST LO parton distributions in a proton $f_i^p$~\cite{MRSTLO} and their similar study in the pion $f_i^{\pi^-}$~\cite{MRSTpion}. The parton distributions in the neutron $f_i^n$ as well as shadowing corrections $f_i^{n / A} / f_i^n$ are given by the proton (nuclear) distributions with the usual assumptions: $u^p = d^n$, $d^p = u^n$, $\bar{u}^p = \bar{d}^n$, $\bar{d}^p = \bar{u}^n$, and $\bar{s}^p = \bar{s}^n$.

The theoretical calculations were then confronted separately with two sets of data. First, the E866/NuSea collaboration reported recently on high-statistics measurements of Drell-Yan dimuon production in proton-nucleus (Be, Fe, W) collisions using the 800 GeV proton beam at Fermilab~\cite{e866}. They extracted both production ratios $R({\rm Fe}/{\rm Be}, x_1)$ and $R({\rm W}/{\rm Be}, x_1)$ over a large kinematic acceptance ($0.28 < x_1 < 0.84$) and on the $4.0 < M < 8.4$~GeV mass range. The second data set corresponds to older measurements from the NA3 collaboration of DY production in pion induced reactions at beam energies $E_{\pi^-} = 150$~GeV and $E_{\pi^-} = 280$~GeV~\cite{na3}. Although the statistics are somehow limited, the ratio of production off hydrogen over platinum targets, $R^{\pi^-}(p / {\rm Pt})$, was extracted at both energies and up to $x_1 \approx 0.9$ in the mass range $4.1 < M < 8.5$~GeV.

%%%%%%%%%%%%%%%%%%%%%%%%%%%%%%%%%%%%%%%%%%%%%%%%%%%%%%%%%%%%%%%%%%%%

\section{Results}\label{sect:results}

\subsubsection*{E866/NuSea data}\label{subsect:results_e866}

Let us first present the results from the fit to the Fermilab data which consist of both ratios $R({\rm Fe}/{\rm Be}, x_1)$ and $R({\rm W}/{\rm Be}, x_1)$ (7 data points each) in $p(800\,{\rm GeV})$--$A$ collisions. The agreement between data and theory is summarized in Table~1 where the $\chi^2$ per number of degrees of freedom ($ndf$)\footnote{The number of degrees of freedom is 14 and 15 depending on whether the energy loss coefficient $\delta$ is taken or not as a free parameter.} is given. 

First, it is clear from Table~1 that the E866/NuSea data exhibit significant nuclear effects from the large $\chi^2/ndf = 4.49$ when neither shadowing nor energy loss is taken into account. Even though nuclear shadowing and/or energy loss gives an excellent description ($\chi^2/ndf \approx 0.5$), these data do not allow one to pin down one or the other case from the constant $\chi^2/ndf$ in each separate scenario. Perhaps more interesting is the amount of energy loss required to describe E866/NuSea measurements. When EKS98 shadowing is not included (iii), it turns out that a large $\delta=3.5$~GeV/fm is required which corresponds, using Eq.~(\ref{eq:eloss}), to an energy loss per unit length of $d E / d z=1.75$~GeV/fm in a large nucleus ($L\approx 5$~fm). This result turns out to be close to (although well smaller than) the recently fixed $d E / d z\approx 2.7$~GeV/fm by Johnson et al.~\cite{johnson} from the $x_1$ and $M$ dependence of E772 and E866/NuSea data. On the contrary, no significant energy loss ($\delta = 0.1$~GeV/fm) is found when EKS98 shadowing is included in the calculations~(iv), which confirms previous results of Ref.~\cite{e866}. However, one should keep in mind that E772 Drell-Yan measurements, taken in the same kinematic range as E866/NuSea, have been used to constrain the EKS98 parameterization. This may therefore explain why E866/NuSea data (consistent with E722 results) are well reproduced assuming EKS98 shadowing only ($\chi^2/ndf = 0.51$, (ii)). This lack of consistency thus weakens our confidence in a vanishing energy loss in Fermilab Drell-Yan data. Therefore, quark energy loss appears to strongly depend on the initial assumptions, as illustrated in Figure~1 ({\it left}) where the $\chi^2/ndf$ is plotted as a function of $\delta$. Moreover, Figure~1 shows that the two fitted values are rather well constrained from the deep minima. To quantify the error on the parameter $\delta$, we display between brackets in Table~1 the upper limits $\delta + \Delta \delta$ where the one standard-deviation error $\Delta \delta$ is given by the deviation of $\chi^2$ by one unit from its minimum.

Consequently, E866/NuSea data do not permit conclusions on the very origin of the observed nuclear dependence because of the present uncertainties in the amount of sea quark shadowing. This prevents us from setting constraints on quark energy loss in nuclear matter. Let us now turn to the NA3 DY dimuon production in pion nucleus reactions.
\vspace{0.3cm}
\begin{table}[htbp]
\begin{center}
\begin{tabular}{|c|c|c|c|c|c|}
\hline
 \multicolumn{2}{|c|}{~} & \hspace{0.7cm}(i)\hspace{0.7cm} & \hspace{0.7cm}{\centering (ii)}\hspace{0.7cm} & {\centering (iii)} & {\centering (iv)} \\
\hline 
& $\delta$ (GeV/fm) &{ \centering ---} & {\centering ---} & {\centering 3.5 (4.0)} & {\centering 0.1 (0.6)} \\
\cline{2-6} 
E866/NuSea & $\chi^2 / ndf $ & {\centering 4.49} & {\centering 0.51} & {\centering 0.52} & {\centering 0.54} \\
\hline
& $\delta$ (GeV/fm) & {\centering ---} & {\centering ---} & {\centering 0.3 (0.7)} & {\centering 0.5 (1.0)}\\
\cline{2-6} 
NA3 & $\chi^2 / ndf $ & {\centering 0.38} & {\centering 0.43} & {\centering 0.39} & {\centering 0.38} \\
\hline
& $\delta$ (GeV/fm) &{ \centering ---} & {\centering ---} & {\centering 1.6 (1.8)} & {\centering 0.3 (0.7)} \\
\cline{2-6}
both & $\chi^2 / ndf $ & {\centering 2.32} & {\centering 0.48} & {\centering 1.34} & {\centering 0.47} \\
\hline
\end{tabular}
\label{tab:fitdy}
\caption{Results from a fit to E866/NuSea to NA3 data sets for the various ansatz assumed in the calculations (see text). The 1 $\sigma$ upper limits $\delta + \Delta \delta$ are given between brackets.}
\end{center}
\end{table}

\subsubsection*{NA3 data}

Unlike the DY data taken at Fermilab, Drell-Yan production in $\pi^-$ induced reactions at $E_{\pi^-}=150$ and 280~GeV should not be spoiled by large shadowing corrections. The reason for this is twofold. First, the mean momentum fraction $x_2$ probed in the NA3 measurements turns out to be much larger ($0.06 < \langle x_2\rangle < 0.3$) than at Fermilab ($0.02 < \langle x_2\rangle < 0.06$) because of the smaller incident energy. In this intermediate $x_2$ range ($x_2 \sim 0.1$), DY should only be slightly affected by quark (anti)shadowing~\cite{EKS98}. Furthermore, the Drell-Yan process in these reactions is dominated by the annihilation of valence quarks for which shadowing is well constrained from DIS measurements only.

The results from the fit to the ratio $R^{\pi^-}(p/Pt, x_1)$ at $E_{\pi^-}=150$ and 280~GeV (resp. 8 and 9 data points) are displayed in Table~1. First, Table~1 shows that the NA3 data are well accounted for without invoking any nuclear effect ($\chi^2/ndf = 0.38$), as already pointed out in~\cite{na3}. The agreement can be seen in Figure~2 ({\it left}) where the measurements at 150~GeV are compared to the calculations ((i), {\it solid})\footnote{Notice that data are consistent with the expected $R^{\pi^-}({\rm p}/{\rm Pt}) = 2 A / (Z+A) \approx 1.43$ assuming only valence-valence fusion process in Eq.~(\ref{eq:dyxsA}) with $u_v^p(x_2) \approx 2 u_v^n(x_2)$ in this $x_2$ range.}. Furthermore, we note that shadowing effects only marginally affect DY production ((ii), {\it dashed}) as expected. As a consequence, we anticipate that energy loss can be fixed from these data without ambiguity anymore.
\begin{figure}[ht]
\begin{center}
\begin{minipage}{6.2cm}
\includegraphics[width=7.cm]{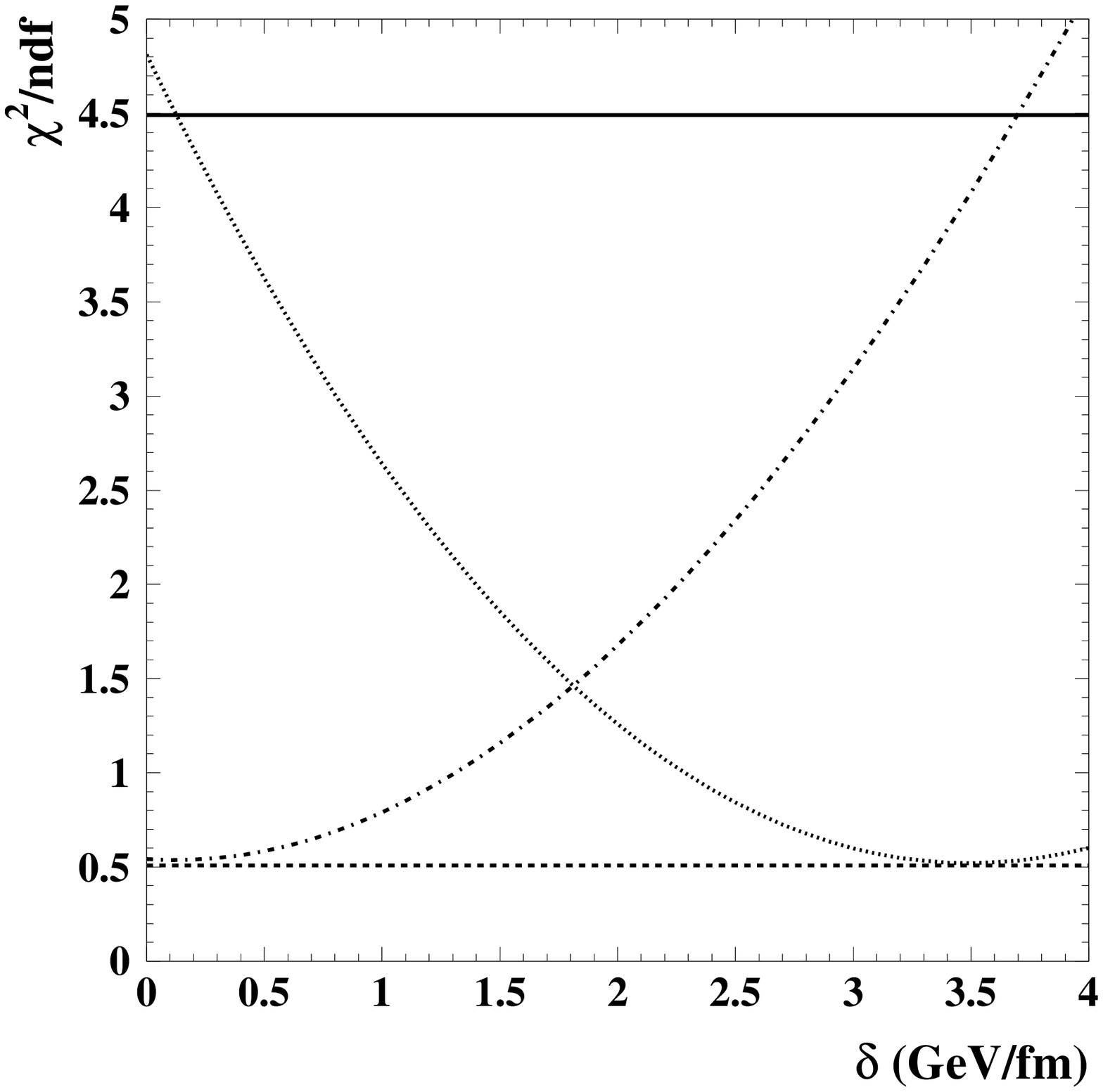}
\end{minipage}
~\hfill
\begin{minipage}{6.2cm}
\includegraphics[width=7.cm]{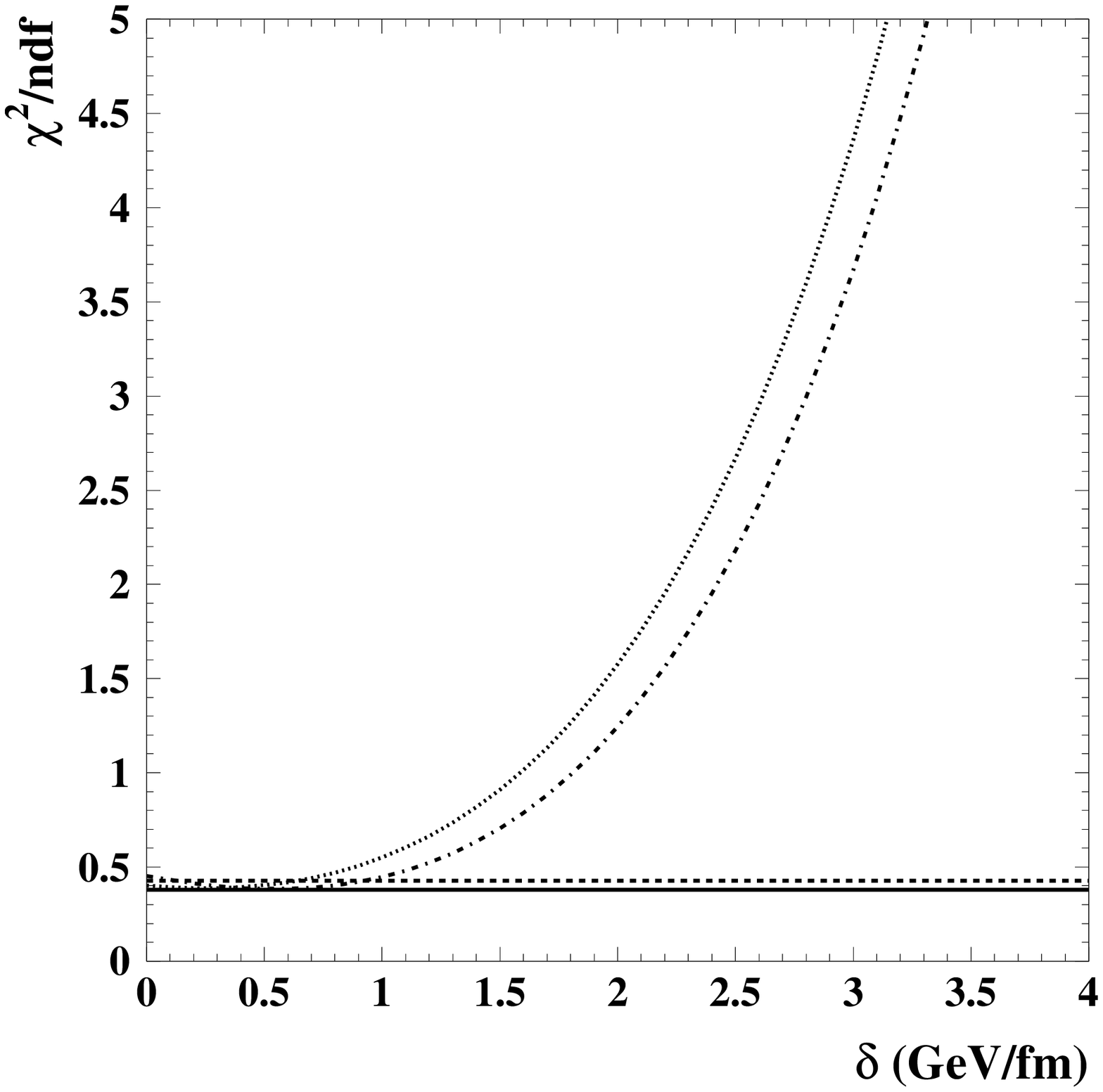}
\end{minipage}
\label{fig:chi2}
\caption{$\chi^2/ndf$ between E866/NuSea ({\it left}) and NA3 ({\it right}) data and theoretical calculations without ((iii), {\it dotted}) and  with ((iv), {\it dash-dotted}) shadowing corrections as a function of the energy loss coefficient $\delta$. The $\chi^2/ndf$ for vanishing energy loss ($\delta = 0$) is also shown without ((i), {\it solid}) and with ((ii), {\it dashed}) shadowing contribution.}
\end{center}
\end{figure}

Within both ansatz (iii) and (iv), NA3 data reveal that quark energy loss is small, from $\delta = 0.3$~GeV/fm up to 0.5~GeV/fm when shadowing is taken into account. What is more, the small one standard-deviation upper limits (0.7 and 1.0~GeV/fm) clearly indicate that DY measurements at SPS --- despite the large error bars --- put stringent constraints on the maximal quark energy loss in nuclear matter. The strong disagreement between NA3 data and theory for too large energy loss coefficients $\delta$ is shown in Figure~1 ({\it right}) where the $\chi^2/ndf$ is plotted. In particular, they allow to exclude the huge energy loss $\delta = 3.5\pm 0.5$~GeV/fm extracted from the E866/NuSea DY data assuming no shadowing effects (cf. Table~1, (iii)). To get a feeling for the origin of such tight constraints, we plot in Figure~2 ({\it right}) the theoretical predictions for $\delta = 0$ ({\it solid}), 1.5 ({\it dashed}) and 3 ({\it dotted}) GeV/fm in comparison to the 150~GeV data. There, we see that the effects of quark energy loss become significant at large $x_1$, leading to a fast increase of the ratio $R^{\pi^-}({\rm p}/{\rm Pt}, x_1)$ in contradistinction to the trend of the data. Even though the increase of $\chi^2/ndf$ comes mainly from the region of large $x_1$ measurements, it is worth noting that data above $x_1 \ge 0.6$ cannot be accommodated with, say, $\delta = 3$~GeV/fm. From the combined fit to NA3 data with (iv) and without (iii) shadowing, we found the energy loss coefficient to be $\delta~=0.4~\pm~0.3\,\,{\rm GeV/fm}$ for fast quarks in nuclei. We shall discuss this result in the next section.

Finally, a global analysis of the E866/NuSea and NA3 overall data has been performed. As expected, Table~1 not only indicates that data show a significant nuclear dependence ($\chi^2/ndf = 2.32$) but clearly demonstrate that the energy loss mechanism alone cannot account for it ($\chi^2/ndf = 1.34$). Hence, this is a clue that shadowing is at the origin of the nuclear dependence observed at Fermilab.

\section{Discussion}\label{sect:discussion}

The result of our analysis is that NA3 pion-nucleus data proved more effective than the Fermilab precise measurements to constrain the amount of quark energy loss in matter which turns out to be small: $\delta=0.4\pm~0.3$~GeV/fm. Before discussing this result and comparing it with previous studies, we would like to comment on limitations and uncertainties in this approach.

First, the tight constraints on maximal quark energy loss arise mainly from the large $x_1$ measurements at 150~GeV (see Figure~2) as already discussed. To investigate the sensitivity of these data points on our final result, a similar procedure has been performed {\it removing}, respectively, the largest ($x_1 = 0.95$) and the three largest valued ($x_1 \ge 0.7$) $x_1$ data points, leading to $\delta = 0.5 \pm 0.4$~GeV/fm and $\delta = 0.7 \pm 0.5$~GeV/fm. Surprisingly, quark energy loss remains well constrained, although a bit larger, from the whole NA3 measurements. Furthermore, we have checked that the theoretical calculations depend only marginally on a specific choice for the proton parton distributions in Eq.~(\ref{eq:dyxsA_general}). Indeed, similar results were found using either the leading order GRV LO~\cite{GRV} and CTEQ5L~\cite{CTEQ} or the next-to-leading order MRST~\cite{MRST} parton densities. Turning to the pion sector, it is regrettable that quark distributions are much less constrained from the limited available data. Let $\bar{u}_v^{\pi^-} \sim (1-x)^\eta$ be the valence quark distribution in the pion at large $x$. In Ref.~\cite{MRSTpion}, the analysis of NA10 and E615 data lead respectively to $\eta = 1.08 \pm 0.02$ and $\eta = 1.15 \pm 0.02$. Nevertheless, going from $\eta = 1.06$ to $\eta = 1.17$ only affected our final result (assuming $\eta = 1.11$) by a few percent at most. On the theoretical side, the Drell-Yan mechanism has been computed in the QCD-improved parton model to leading order in the coupling, assuming the NLO corrections to vanish in the production ratio $R$. This ansatz is justified as long as nuclear effects involved in both LO and NLO processes remain identical, which may not be true in general. Consider for instance the Compton scattering $q g \to q \gamma^*$ process. The larger gluon energy loss ($C_R = 3$ in Eq.~(\ref{eq:de})) would lead to an even smaller quark energy loss estimate\footnote{On top of a different gluon energy loss, antishadowing may be more pronounced for gluons than for valence quarks in the $x_2 \sim 0.1$ window whereas small $x_2$ shadowing in the gluon and the sea quark channel should be quite similar~\cite{EKS98}.}. However, a complete calculation including all NLO processes would certainly be needed. Concerning the computation of quark energy loss, the distribution $D(\epsilon)$ has been calculated in the soft gluon limit $\omega_c \ll E_q$. It has been {\it a posteriori} checked that, with $\delta = 0.4$~GeV/fm, this relation is fulfilled for quark energy down to $E_q \approx 50$~GeV (i.e., the smallest quark energy in the NA3 data). Furthermore, the Glauber approximation (i.e., that multiple successive quark-nucleon scatterings are independent) on which the BDMPS framework relies is only relevant for highly energetic quarks and should break down at very large energies when shadowing effects become large~\cite{BDMPS}. Therefore, we can assume that the use of a Glauber based approach to describe moderate and large $x_1$ data at SPS energy (not too small $x_2$) is sufficiently accurate. As for shadowing corrections, we believe the factorized form (\ref{eq:paramEKS98}) assumed in Ref.~\cite{EKS98} to be meaningful when $x_2$ is not too small, where DY production has been measured ($x_2 > 10^{-2}$). What is more, the use of the EKS98 parameterization which is well constrained from both DIS and DY data ---although {\it not} fitted to the above NA3 measurements--- appears to be fully justified. 

\begin{figure}[ht]
\begin{center}
\begin{minipage}{6.2cm}
\includegraphics[width=7.0cm]{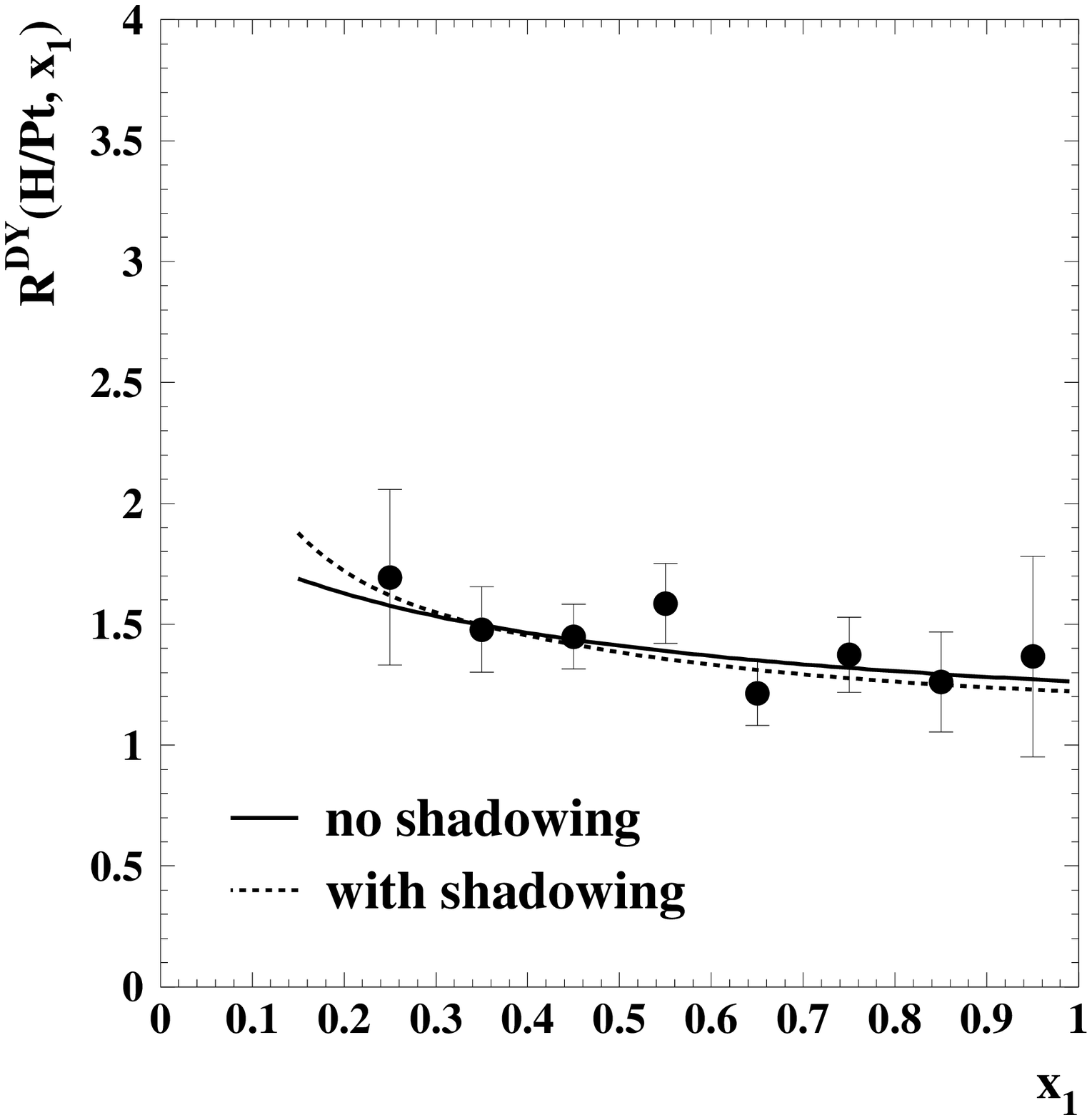}
\end{minipage}
\hfill
\begin{minipage}{6.2cm}
\includegraphics[width=7.0cm]{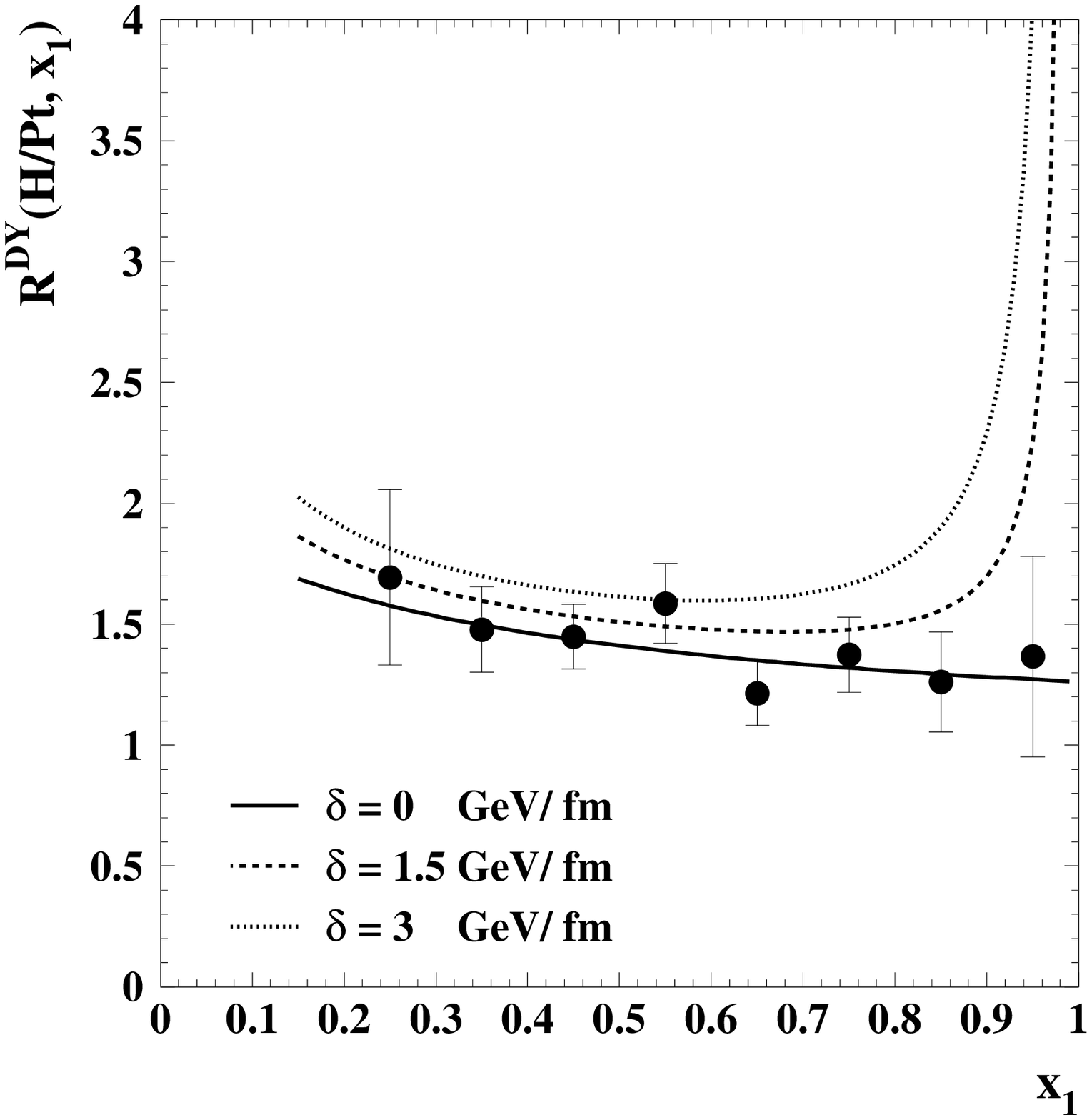}
\end{minipage}
\caption{NA3 ratio $R^{\pi^-}(p/{\rm Pt})$ of Drell-Yan dimuon production versus $x_1$ in $\pi^-$(150 GeV)-A collisions. Calculations assuming shadowing ($left$) and energy loss ($right$) effects are compared to the data.}
\label{fig:na3data}
\end{center}
\end{figure}

We pointed out that several attempts to extract quark energy loss in nuclear matter from the E772 and E866/NuSea Drell-Yan data have recently been made~\cite{e866,johnson}. In Ref.~\cite{e866}, large energy loss effects have been claimed to be ruled out using E866/NuSea data corrected for shadowing and assuming various models for parton energy loss. Using the BDMPS approach, the one standard upper limit was found to be $\delta+\Delta\delta = 0.46$~GeV/fm, comparable to what is quoted here. Nevertheless, as mentioned repeatedly, their result should be taken with care since the shadowing parameterization is partially fitted to E772 data. To avoid such an inconsistency, a new analysis of both E722 and E866/NuSea results came out later in which nuclear shadowing was estimated theoretically~\cite{johnson}. Because of the small shadowing effects in their calculation, a huge quark energy loss $dE/dz = 2.7\pm0.4\pm0.5$~GeV/fm proved necessary to describe the overall Fermilab data set, i.e., rather close to (although somewhat larger than) our estimate $dE/dz = 1.75 \pm 0.25$~GeV/fm (with $L = 5$~fm) from the E866/NuSea analysis assuming no shadowing (Table~1, (iii)). Such a large number cannot be understood as coming from radiative energy loss only~\cite{BDMPS}. It has been attributed in Ref.~\cite{johnson} to the interplay between the effects of gluon radiation together with the energy loss due to the string tension. The result quoted in Ref.~\cite{johnson} is likely to be ruled out from a comparison with the NA3 measurements. In particular, our present work does not show any evidence for quark energy loss coming from the tension of the string stretched from the beam parton to the nucleus. Finally, we would like to stress that the {\it distribution} in the induced energy loss $D(\epsilon)$ has been employed in the present study whereas the previous analysis~\cite{e866,johnson} modeled the quenching by shifting the projectile quark energy by the {\it mean} energy loss $\Delta E$~\footnote{This would correspond to take $D(\epsilon) = \delta(\epsilon - \Delta E)$ in Eq.~(\ref{eq:dyxsA_general})}. As recently pointed out by Baier et al., this standard modeling of the suppression is inadequate when the cross section sharply falls down with $x_1$~\cite{BDMS_quench}. In particular, it is argued that the {\it typical} quark energy loss, i.e., the loss that really contributes to the quenching, proves much smaller than the {\it mean} energy loss $\Delta E$.

It is interesting to note that $- dE/dz = (0.4\pm0.3~{\rm GeV/fm})\,(L / 10\,{\rm fm})$ revealed from the NA3 data proves in excellent agreement with the BDMPS prediction\footnote{The prediction in Ref.~\cite{BDMPS} has to be increased by a factor of two from elastic corrections originally not taken into account~\cite{BDMS}.} $- dE/dz = (0.4~{\rm GeV/fm})\,(L / 10\,{\rm fm})$ in Ref.~\cite{BDMPS}. However, this apparent agreement is coincidental because of both the quoted error bars and the rough estimate in~\cite{BDMPS}. In particular, let us stress that the NA3 data are compatible with zero energy loss (Table~1). Nevertheless, the ``agreement'' is a hint that the origin of the fitted energy loss is radiative. This allows us to extract the ``transport coefficient'' for cold QCD matter $\hat{q} = 0.24 \pm 0.18$~GeV/fm$^2$, which corresponds to a $p_\perp$ broadening $d p_{\perp}^2 / dz = 0.021 \pm 0.016$~GeV$^2$/fm in agreement with E772 results~\cite{e772}. Unfortunately, such a quantity has not been measured by the NA3 collaboration to check the relation between radiative energy loss and $p_\perp$ broadening.

To summarize, a LO analysis of Drell-Yan data in $p$--$A$ and $\pi^-$--$A$ reactions at Fermilab and SPS energies has been carried out. The aim was to set tight constraints on quark energy loss in nuclear matter in a (as much as possible) model-independent way. For this, multiple fits to the data have been performed under various assumptions as for the nuclear effects. At Fermilab energy, Drell-Yan measurements probe a small $x_2$ range where the amount of sea quark shadowing is only poorly known from DIS data. Consequently, we were unable to explore the origin of the nuclear dependence seen in the data and thus to extract unambiguously quark energy loss. On the other hand, nuclear shadowing of valence quarks in the intermediate $x_2$ range is rather small and well constrained from DIS data. Therefore, DY production in $\pi$--$A$ collisions at SPS energies is {\it only} sensitive to energy loss of the incoming valence antiquark. From NA3 measurements, we give the estimate $- dE/dz = 0.20 \pm 0.15$~GeV/fm for the quark mean energy loss in a Platinum nucleus, in good agreement with the expectation from the BDMPS perturbative approach. While smaller error bars would be needed to fix it more precisely, these data already allow one to rule out a quark mean energy loss much greater than 0.5~GeV/fm. This small radiative energy loss thus gives a hint that most of the Drell-Yan nuclear dependence at Fermilab actually comes from large shadowing corrections (as first assumed in the EKS98 parameterization) which may be even stronger at RHIC and LHC energies.

\section*{Acknowledgements} 

I would like to thank J\"org Aichelin, Pol-Bernard Gossiaux and Thierry Gousset for a very pleasant collaboration. I am also grateful to Evgeni Kolomeitsev, St\'ephane Peign\'e, Carlos Salgado, and Wolfram Weise for useful comments and discussions.

\bibliographystyle{unsrt}

\begin{thebibliography}{99}

\bibitem{GP1}
M.~Gyulassy and M.~Pl\"umer, Phys. Lett. B {\bf 243}, 432 (1990);\\
M.~Gyulassy and X.-N.~Wang, Nucl. Phys. B{\bf 420}, 583 (1994);\\
X.-N.~Wang, M.~Gyulassy, and M.~Pl\"umer, Phys. Rev. D{\bf 51}, 3436 (1995).

\bibitem{Baier1}
R.~Baier, Yu.~L.~Dokshitzer, S.~Peign\'e, and D.~Schiff, Phys. Lett. B {\bf 345}, 277 (1995);\\
R.~Baier, Yu.~L.~Dokshitzer, A.~H. Mueller, S.~Peign\'e, and D.~Schiff, Nucl. Phys. B{\bf 483}, 291 (1997).

\bibitem{Wang}
X.-N.~Wang and M.~Gyulassy, Phys. Rev. Lett. {\bf 68}, 1480 (1992).

\bibitem{BDMPS}
R.~Baier, Yu.~L.~Dokshitzer, A.~H. Mueller, S.~Peign\'e, and D.~Schiff, Nucl. Phys. B{\bf 484}, 265 (1997).

\bibitem{phenix}
K.~Adcox {\it et~al.} (PHENIX), Phys. Rev. Lett. {\bf 88}, 022301 (2002).

\bibitem{e866}
M.~A. Vasiliev {\it et~al.} (E866/NuSea), Phys. Rev. Lett. {\bf 83}, 2304 (1999).

\bibitem{johnson}
M.~B. Johnson {\it et~al.}, Phys. Rev. C{\bf 65}, 025203 (2002).

\bibitem{na3}
J. Badier {\it et al.} (NA3), Phys. Lett. B {\bf 104}, 335 (1981); \\
O.~Callot {\it et al.} (NA3), Contribution to Moriond Workshop on Lepton Pair Production, Les Arcs (1981).

\bibitem{gavin}
S.~Gavin {\it et~al.}, Int. J. Mod. Phys. A {\bf 10}, 2961 (1995).

\bibitem{arneodo}
For a review, see M.~Arneodo, Phys. Rept. {\bf 240}, 301 (1994);\\
G.~Piller and W.~Weise, Phys. Rept. {\bf 330}, 1 (2000) and references therein.

\bibitem{QS}
J.~Qiu and G.~Sterman, preprint hep-ph/0111002 (2001).

\bibitem{EKS98}
K.~J. Eskola, V.~J. Kolhinen, and C.~A. Salgado, Eur. Phys. J. C{\bf 9}, 61 (1999).

\bibitem{e772}
D.~M. Alde {\it et~al.} (E772), Phys. Rev. Lett. {\bf 64}, 2479 (1990).

\bibitem{BDMS_quench}
R.~Baier, Yu.~L.~Dokshitzer, A.~H. Mueller, and D.~Schiff, JHEP {\bf 0109}, 033 (2001).

\bibitem{BDMS}
R.~Baier, Yu.~L.~Dokshitzer, A.~H. Mueller, and D.~Schiff, Nucl. Phys. B{\bf 531}, 403 (1998).

\bibitem{MRSTLO}
A.~D.~Martin, R.~G.~Roberts, W.~J.~Stirling, and R.~S. Thorne, Phys. Lett. B {\bf 443}, 301 (1998).

\bibitem{MRSTpion}
P.~J.~Sutton, A.~D.~Martin, R.~G.~Roberts, and W.~J.~Stirling, Phys. Rev. D{\bf 45}, 2349 (1992).

\bibitem{GRV}
M.~Gl\"uck, E.~Reya, and A.~Vogt, Eur. Phys. J. C{\bf 5}, 461 (1998).

\bibitem{CTEQ}
H.~L.~Lai {\it et~al.} (CTEQ), Eur. Phys. J. C{\bf 12}, 375 (2000).

\bibitem{MRST}
A.~D.~Martin, R.~G.~Roberts, W.~J.~Stirling, and R.~S. Thorne, Eur. Phys. J. C{\bf 4}, 463 (1998).

\end{thebibliography}

\end{document}